\documentclass{raa}
\usepackage{graphicx,times}
\usepackage{natbib}
\usepackage{amssymb,amsmath}
\bibpunct{(}{)}{;}{a}{}{,}

\usepackage[a4paper=true,pagebackref=true]{hyperref}
\hypersetup{pdftitle = The title of my PDF, pdfauthor = My name, pdfsubject= The subject, pdfkeywords = keyword1 keyword2 keyword3}
\hypersetup{colorlinks = true, linkcolor = green, anchorcolor = red, citecolor = blue, filecolor = red, pagecolor = red, urlcolor = red}

\begin{document}

   \title{A Strange Star Scenario for the Formation of Eccentric Millisecond Pulsar PSR J1946+3417$^*$
\footnotetext{\small $*$ Supported by Key laboratory of Modern Astronomy and Astrophysics (Nanjing University), Ministry of Education, China.}
}

 \volnopage{ {\bf 20XX} Vol.\ {\bf X} No. {\bf XX}, 000--000}
   \setcounter{page}{1}

   \author{L. Jiang\inst{1, 2, 3}, Na Wang \inst{1}, Wen-Cong Chen\inst{2, 3}, Wei-Min Liu\inst{3}, Chun-wei Leng\inst{3}, Jian-Ping Yuan\inst{1}, Xiang-Li Qian\inst{4} }

   \institute{ Xinjiang Astronomical Observatory, CAS, Urumqi, Xinjiang 830011, China
        \and
             School of Science, Qingdao University of Technology, Qingdao 266525, China\\
	\and
             School of Physics and Electrical Information, Shangqiu Normal University, Shangqiu 476000, China\\
         \and
              Department of Intelligent Engineering, Shandong Management University, Jinan 250357, China\\
               {\it Wang: na.wang@xao.ac.cn; Chen: chenwc@pku.edu.cn}\\
\vs \no
   {\small Received 20XX Month Day; accepted 20XX Month Day}
}

\abstract{PSR J$1946+3417$ is a millisecond pulsar (MSP) with a spin period $P\simeq3.17\rm~ms$.
Harbored in a binary with an orbital period $P_{\rm b}\simeq27$ days, the MSP is accompanied by a white dwarf (WD).
The masses of the MSP and the WD were determined to be $1.83\rm~M_\odot$ and $0.266\rm~M_\odot$, respectively.
Specially, its orbital eccentricity is $e\simeq0.134$, which is challenging the recycling model of MSPs.
Assuming that the neutron star in a binary may collapse to a strange star when its mass reaches a critical limit,
we propose a phase transition (PT) scenario to account for the origin of the system.
The sudden mass loss and the kick induced by asymmetric collapse during the PT may result in the orbital eccentricity.
If the PT event takes place after the mass transfer ceases, the eccentric orbit can not be re-circularized in the Hubble time.
Aiming at the masses of both components, we simulate the evolution of the progenitor of PSR J$1946+3417$ via \texttt{MESA}.
The simulations show that a NS / main sequence star binary with initial masses of $1.4+1.6\rm~M_\odot$ in an initial orbit of 2.59 days will evolve into a binary consisting of a $2.0\rm~M_\odot$ MSP and a $0.27\rm~M_\odot$ WD in an orbit of $\sim21.5$ days.
Assuming that the gravitational mass loss fraction during PT is $10\%$, we simulate the effect of PT
via the kick program of \texttt{BSE} with a velocity of $\sigma_{\rm PT}=60~{\rm km~s}^{-1}$.
The results show that the PT scenario can reproduce the observed orbital period and eccentricity with higher probability then other values.
}

   \authorrunning{L. Jiang, N. Wang et al. }            
   \titlerunning{Strange star scenario for PSR J1946+3417}  
   \maketitle

%
\section{Introduction}           
\label{sect:intro}

Pulsars with low surface magnetic fields ($B\simeq10^8-10^9{\rm ~G}$) and short spin periods ($P\leq20{\rm ~ms}$)
are known as millisecond pulsars (MSPs).
They are neutron stars (NSs) which were spun up to millisecond period in low-mass X-ray binaries (LMXBs)
by the accretion of mass and angular momentum \citep{man04, lor08}.
This is the widely accepted standard recycling model \citep{alp82, rad82, bha91}.
Harbored in pre-LMXBs with initial orbital periods longer than the so-called bifurcation period \citep{pyl89},
the donor stars always lose their hydrogen envelopes and evolve into He white dwarfs (WDs).
Considering the tidal interaction with a long timescale during the mass transfer,
MSPs are expected to be found in binaries with circularized-orbits \citep{phi92}
unless they have ever experienced some dynamical processes in dense globular clusters \citep{ver87, ver88}.

The discovery of several eccentric MSP (eMSP) binaries presented a challenge to recycling model.
PSR J1903$+$0327 (hereafter J1903) is the first reported eMSP located in the Galactic field.
Accompanied by a G-type main sequence (MS) star, the eccentricity of this source is $e\simeq0.44$ \citep{cha08}.
\citet{liu09} suggested that it is a newborn NS which is experiencing an accretion from the supernova fallback disk,
while another evolutionary channel suggests that it originates from a hierarchical triple system \citep{fre11, por11, pij12}.
PSR J1618$-$3921 (hereafter J1618) is the second reported Galactic field eMSP with $e\simeq0.027$ \citep{bai10}.
Its spin period is $P\simeq12{\rm~ms}$, and the orbital period is $P_{\rm b}\simeq22.7{\rm~days}$.
According to its mass function, the companion star should be a He WD \citep{oct18}.

Subsequently, four Galactic field eMSPs with similar orbital properties have been discovered.
They are PSRs J2234$+$0611 \citep[hereafter J2234,][]{den13}, J1946$+$3417 \citep[hereafter J1946,][]{bar13},
J1950$+$2414 \citep[hereafter J1950,][]{kni15}, and J0955$-$6150 \citep[hereafter J0955,][]{cam15}.
As listed in Table 1, they share some similar properties, i.e., (1) orbital eccentricities vary from 0.08 to 0.14,
(2) accompanied by He WDs with masses from 0.2 to 0.3$~M_{\odot}$, (3) orbital periods lie in the range from 22 to 32 days,
(4) their short spin periods ($P\leq 5.0{\rm~ms}$) which indicate long-lasting mass transfer episodes during their LMXB phase. Their strong similarities are not expected to result from the disruption of triple systems like J1903, since such a chaotic process should produce a wide range of orbital eccentricities and orbital periods.

\begin{table*}
\begin{center}
\footnotesize
\caption{Some observed parameters of six known Galactic field eMSPs
\\(See http://www.atnf.csiro.au/research/pulsar/psrcat, \citet{man05})}
\begin{tabular*}{\textwidth}{@{}cccccccccl@{}}
\hline\hline\noalign{\smallskip}
\footnotesize
PSR & $P~{(\rm ms)}$ & $\dot{P}~{(\rm s~s^{-1})}$ & $P_{\rm b}~({\rm~days})$ & $e$  & $M_{\rm 1}~({\rm M}_{\odot})$  & $M_{\rm 2}~({\rm M}_{\odot})$ & Companion type & $\tau(\rm ~Gyr)^{\ast}$ & Ref.\\
\hline\noalign{\smallskip}
J1903  & $2.15$      & $1.88\times10^{-20}$   &95.2  &   0.437   &  -  &   $1.08^{\star}$&MS  &$1.81$ & \citet{fre11}\\ \hline
J1618   & $11.99$    & $5.41\times10^{-20}$     &22.7   &   0.027 &  -   &   $0.20^{\star}$&He WD  &$3.51$ & \citet{bai10}\\
J2234  & $3.58$      & $1.2\times10^{-20}$       &32.0   &  0.129   &  $1.353$  &  $0.298$ &He WD &$4.72$& \citet{sto19}\\
J1950  & $4.30$      & $1.88\times10^{-20}$    &22.2    &  0.080   &  $1.496$  &    $0.28$  &He WD& $3.63$& \citet{zhu19}\\
J1946  & $3.17$      & $3.12\times10^{-21}$    &27.02  & 0.1345  &  $1.828$ &   $0.2656$&He WD &$16.1$& \citet{bar17}\\
J0955  & $2.00$     & $-$                    &24.6    &  0.11      & -         &  $\geq0.21$ & He WD &-& \citet{cam15}\\
\hline\noalign{\smallskip}
\end{tabular*}\\
\end{center}
\tablecomments{0.9\textwidth}{\small{$\ast$ $\tau\equiv P/2\dot{P}$, the characteristic age; $\star$: No further measurement available, here is the median masses derived from the mass function assuming a random distribution of orbital inclinations and a pulsar mass of 1.35$\rm~M_{\odot}$.}}
\end{table*}

At present, three evolutionary channels forming eMSPs were proposed. First, the accretion-induced collapse \citep[AIC,][]{nom79, miy80, taa86, can90, nom91} of massive ONe WDs may produce MSPs \citep[for review, see][]{wan20}.
\citet{fre14} suggested that these eMSPs were formed by the rotation delayed AIC (RD-AIC) of massive ONe WDs with masses of $\sim1.2{\rm~M_{\odot}}$. In a binary consisting of a massive ONe WD and a MS companion,
the evolving MS star will transfer hydrogen-rich material onto the WD  when it overflows its Roche lobe.
For a rapidly rotating WD, its mass can exceed the Chandrasekhar limit without collapsing to an NS \citep{yoo04}.
When the mass transfer ceased, the WD will spin-down. If the centrifugal forces cannot sustain the hydrostatic equilibrium, the WD will collapse. The sudden released gravitational binding energy during the collapse will imposes eccentricity on the circular orbit. Since the re-circularization timescale are generally much longer than Hubble time \citep{zah77, hut81}, such MSP/He WD binaries will be eMSP binaries. Second, \citet{ant14} proposed a circumbinary (CB) disk scenario.
He suggest that the dynamical interaction between the binary and a CB disk may cause the eccentricity.
In this scenario, the CB disk originates from the escaping material from the donor star during hydrogen-shell flash shortly before the WD cooling phase.
Adopting the linear perturbation theory given by \citet{der13}, his calculation show that  a CB disk with a fine-tuned lifetime as long as $\sim10^5$ years and a mass around $\sim10^{-4}{\rm~M_{\odot}}$ could result in an eccentricity of $0.01\leq e \leq 0.15$ for post-LMXBs with orbital periods between 15 and 50 days.
Third, considering that phase transition (PT) from NSs to strange stars (SSs) may occur when the core density of accreting NSs in LMXBs reaches the critical density for quark de-confinement, \citet{jia15} (hereafter Paper 1) argued the NS-SS PT scenario. Similar to the RD-AIC model, the calculation in Paper 1 shows that the sudden gravitational mass loss of the NS during PT can produce the observed eccentricities.

Recently, accurate mass measurements of some eMSPs were reported, which may provide some clues on their formation channel. As shown in Table 1, the measurement of J2234 yields $M_{\rm 1}\simeq1.35\rm~M_{\odot}$ for the MSP
and $M_{\rm 2}\simeq0.30\rm~M_{\odot}$ for its companion \citep{sto19}.
Based on the timing observations of J1950 from the data of Arecibo ALFA pulsar survey,
\citet{zhu19} derived $M_{\rm 1}\simeq1.50\rm~M_{\odot}$, and
$M_{\rm 2}\simeq0.28\rm~M_{\odot}$ for the MSP and the WD, respectively.
It seems that the mass measurements of both J1950 and J2234 prefer to the RD-AIC scenario.
However, this scenario tends to produce a low-mass MSP. Even if the mass increase of WD due to differential rotation are considered, a specific accretion rate is also difficult to satisfy \citep{yoo04}. Therefore, the RD-AIC scenario is difficult to account for the formation of J1946 with a high MSP mass ($M_{\rm 1}\simeq1.83\rm~M_{\odot}$) \citep{bar17}.
Furthermore, CB disks surrounding the binary MSPs can result in the formation of eMSPs with high NS masses,
whereas the hydrogen-shell flashes generally exist in the final evolutionary stages of LMXBs, which is contrary to the rarity of eMSPs in the binary MSPs population. As a peculiar eMSP, the formation of PSR J1946 is difficult to understand by the RD-AIC or CB disk scenario. Employing a detailed binary evolution model, in this work we attempt to diagnose whether the NS-SS PT scenario can be responsible for the formation of PSR J1946. 

This paper is arranged as follows.
The NS-SS PT scenario is deiscussed in section 2, , whereas the stellar evolution code and the spin evolution of NS are described in section 3. The simulated results and the influence of PT  kick on the evolution of post-LMXBs are presented in sections 4 and 5, respectively. Finally, we give a brief summary in section 6.

\section{Strange Star Scenario}\label{sec:2}
Considering that the strange quark matter is most stable, the concept of SS
was proposed \citep{ito70, bod71, far84, wit84, alc86, hae86, xu03, lai09, liu12}, i. e. some pulsars may be SSs instead of NSs.
Believing that both NSs and SSs coexist in the Universe, it was argued that NS-SS PT may occur when the central density of an NS rises above the quark de-confinement density \citep{ber03, bom04, bom08, zhu13, bom16, dra16, bha17, wik17, hou18, alv19}.

Studying the equation of state (EoS) of neutron-rich matter,
\citet{sta06} propose that the critical density for quark de-confinement is about five times of the nuclear saturation density.
For a rapidly spinning NS, the centrifugal force might reduce the central density and increase the maximum mass.
With spin period $P$, this can be expressed as \citep{har70,bay71}:
\begin{equation}
M_{\rm c}(P)=M_{\rm c}(0)+\delta M(P_{\rm min} /P)^2,
\end{equation}
where $P_{\rm min}$ is the minimum spin period (in this work, we take $P_{\rm min}=1 {\rm ~ms}$ ), $M_{\rm c}(0)$ is the maximum mass of the non-rotating NS, $\delta M$ denotes the rotation induced maximum mass increase.
\citet{las96} found that for a rigid rotating NS, $\delta M/M_{\rm c}(0)\sim20\%$ while
\citet{mor04} and \citet{hae07} derived that this ratio is about $50\%$ for a differentially rotating NS.

In current work, we take $M_{\rm c}(0)=1.8{\rm ~M}_\odot$ \citep{akm98}, and
$\delta M=0.4~{\rm ~M}_\odot$ \citep{las96}. PT process is assumed to take place when the mass of the NS, $M_{\rm NS}$, exceeds its maximum mass at a spin period of $P$, $M_{\rm c}(P)$. It can occur in the spin-up stage induced by the accretion, or in the spin-down stage of NS after the accretion ceased.
The latter is defined as delayed PT, which can be applied to J1946.

In the recycling stage, the NS would be spun up by accreted material from the donor star.
\citet{che00} derived an expression for the spin period evolution of the NS:
\begin{equation}
\begin{split}
\ P={\rm max}[1.1\left(\frac{M_{\rm NS}-M_{\rm NS, i}}{M_{\odot}}\right)^{-1}R_6^{-5/14}I_{45} \left(\frac{M_{\rm NS}}{M_{\odot}}\right)^{-1/2},\\
1.1\left(\frac{M_{\rm NS}}{M_{\odot}}\right)^{-1/2}R_6^{17/14}]~~{\rm ms},
\end{split}
\end{equation}
where $R_6$ is the radius of the NS in units of $10^6{\rm ~cm}$, $I_{45}$ is the moment of inertia of the NS in units of $10^{45}{\rm~g}{\rm ~cm}^{2}$, ($R_6=I_{45}=1$ in this work), and $M_{\rm NS, i}$ is the initial mass of the NS.

The difference between the gravitational masses of NS and SS with the same baryon number is also widely studied \citep{bom00, dra07, mar17}. It is generally thought $\Delta{M}=M_{\rm NS}-M_{\rm SS}\approx 0.15~{\rm ~M_\odot}$ for NS with a mass of $M_{\rm NS}\simeq1.5~{\rm ~M_\odot}$ \footnote
{A lower value may also possible. For instance, \citet{sch02} suggest
that the difference of the gravitational masses between NS and hyperon star is $\simeq0.03\rm {~M_\odot}$.}, i. e. the mass loss ratio $\Delta{M}/M_{\rm NS}$ during PT is about 10\%.
Following their researches, in this work, we take the mass of SS $M_{\rm SS}=0.9M_{\rm NS}$.
Assuming J1946 is a SS formed via NS-SS PT, the mass of the NS before PT can be derived to be $2.00-2.06\rm~M_{\odot}$ according to the current mass $M_{\rm SS}=1.828(22)\rm~M_{\odot}$ \citep{bar17}.

The detailed PT process from baryons to quarks is not well understood.
\citet{oli87} and \citet{hor88} believed that the PT will last around $10^8{\rm ~yr}$ as a gradual process,
while many researchers argued that the process might be a detonation mode with timescale from several milliseconds to $\sim10$ seconds \citep{che96, ouy02, shu17, pra18, zha18, mar19, ouy12, ouy19, sin20}.
Their studies show that the energy released during PT is compatible with a core collapse supernova (CCSN).
We consider that the PT takes place in the core of NS like CCSN,
and a kick velocity $V_{\rm k}$ is imparted to the new born SS due to a spherical asymmetric collapse\footnote{The disruption of spherical symmetry might originate from the fast rotation of NSs or the dipole magnetic field. However, the detailed process is currently poorly understood.}.
Since the mass transfer timescale is of order of $10^8\rm yr$, the binary orbit before PT is assumed to be circular.
Setting the positional angle of $V_{\rm k}$ with respect to the pre-PT orbital plane as $\phi$,
and the angle between $V_{\rm k}$ and the pre-PT orbital velocity $V_{\rm 0}$ as $\theta$,
then the ratio between the semi-major axes before and after PT is \citep[]{hil83, dew03, sha16}
\begin{equation}
\frac{a_0}{a}=2-\frac{M_{\rm T}}{M_{\rm T}-\Delta{M}}(1+\nu+2\nu~{\rm cos}~\theta),
\end{equation}
where $\nu=V_{\rm k}/V_{\rm 0}$, $V_{\rm 0}=(2\pi GM_{\rm T}/P_{\rm b, 0})^{1/3}$,
$P_{\rm b, 0}$ and $M_{\rm T}$  are the orbital period and the total mass of the binary before PT, respectively.
Because of the influence of sudden mass loss and the kick, the eccentricity after PT satisfies \citep[]{hil83, dew03, sha16}
\begin{equation}
\begin{split}
e^2=1-\frac{a_{\rm 0}M_{\rm T}}{a({M_{\rm T}-\Delta{M})}}[1+ 2\nu~{\rm cos}~\theta  \\
+ \nu^2({\rm cos}^2\theta + {\rm sin}^2\theta~{\rm sin}^2\phi)].
\end{split}
\end{equation}

The kick velocity distribution during PT is described as a Maxwellian distribution with one-dimensional rms $\sigma_{\rm PT}$. By analysing the proper motions of 233 pulsars \citet{hob05} obtained a mean speed of $54(6) {\rm km s}^{-1}$ for recycled pulsars. Employing the rapid binary population synthesis (\texttt{BPS}) code \citep{hur00,hur02} and a weak kick velocity of $\sigma_{\rm PT} = 60\rm~km~s^{-1}$,
\citet{jiang20} (hereafter Paper 2) studied the formation of isolated MSPs via NS-SS PT \citep[a similar work is performed by][]{nur19}.
Paper 2 shows that, almost all the PT processes take place at the Roche Lobe overflow (RLOF) stage.
However, it is also possible to take place after mass transfer ceases, and induce the formation of eccentric binary MSPs like J1946.

\section{Simulation Description}
\subsection{Binary Evolution}
To obtain the orbital properties of J1946 before PT,
we simulated the evolution of its progenitor via \textsf{MESAbinary} in \texttt{MESA} (version $\rm r-9575$) \citep{pax11, pax13, pax15}. The starting point of the simulation is a circular binary system containing a primary NS 
and a zero-age MS companion star 
with a solar composition ($X= 0.70$, $Y= 0.28$, and, $Z= 0.02$).

During the mass transfer, we adopt the accretion efficiency $f=1-\alpha-\beta-\delta$
described by \citet{tau06}, where $\alpha$ and $\beta$ are the fraction of mass lost from the vicinity of the donor and the NS, respectively, while $\delta$ is the fraction of mass lost as coplanar toroid circling the binary.
In current work, $\alpha=\delta=0$ is adopted.
In addition, a fixed Eddington accretion rate of the NS, $\dot{M}_{\rm Edd}=1.8\times10^{-8}{\rm ~M_\odot~yr^{-1}}$, is also considered, i.e. $\dot{M}_{\rm NS}={\rm min}(-f\dot{M}_{\rm d}, \dot{M}_{\rm Edd})$,
which may cause some more mass loss from the vicinity of the NS.
Besides the angular momentum loss due to mass loss as fast winds \citep{pax15, den21},
effects of gravitational radiation \citep{lan59, fau71} and magnetic braking \citep[with $\gamma= 3.0$,][]{rap83, pax15, den20} are also considered.

\subsection{Spin Evolution}
For a delayed PT pulsar, its spin evolution can be divided into three stages.
The first one is the spin-up process (see also Eq. 2) during the recycling stage.
The second one is the spin-down stage from the endpoint of RLOF(the stellar age is $t_1$) to the beginning of the NS-SS PT, in which the timescale is denoted by$t_{2}$ (it is also defined as the delay time).
The third stage is the spin-down stage after PT (as a SS), denoting with a timescale $t_{3}$.
Due to the uncertainties of EOS, the changes of spin period and magnetic field caused by PT are ignored.

The MSP is assumed to spin down according to a power law \citep{lyn75},
 i. e. the spin period derivative $\dot{P}=K P^{(2-n)}$,
where $n$ is the braking index, and $K$ is always a constant (indeed, it may also vary due to the evolution of magnetic field \citet{lyn75}).
Taking $R_6=I_{45}=1$, pure magnetic dipole radiation model ($n=3$) with the magnetic inclination angle $\chi=90^{\circ}$ predicts
a spin period derivative of the MSP as follows
\begin{equation}
 \dot{P}=0.98\times10^{-20}P_{\rm ms}^{-1}B_{8}^2\rm ~s~s^{-1},
\end{equation}
where $P_{\rm ms}$ is the spin period in units of 1 ms, $B_{8}$ is the surface magnetic field in units of $10^8\rm ~G$ \citep{woa18}.

Considering both the effect of residual field \citep{din93},
 and long term exponential decay \citep{urp94, gus04} with $t_{\rm D}=10^8{\rm ~yr}$ \citep[]{bra18, xie19}\footnote
 {Smaller or larger values are also derived, for example, $3\times10^6{\rm~yr}$ \citep{gus04}, and
 $1.8\times10^9{\rm ~yr}$ \citep{bra18}. See \citet{sun02, rud10}, for review.},
we assume
\begin{equation}
B={\rm max}[B_{\rm 0}{\rm exp}(-t/t_{\rm D}), ~B_{\rm min}],
\end{equation}
where $B_{\rm 0}$ and $B_{\rm min}$ are the initial value and the minimum of surface magnetic field, respectively.
In this work, we adopt a typical initial magnetic field of MSPs $B_{\rm 0}=10^9\rm~G$ , and $B_{\rm min}=10^8\rm ~G$ following \citet{din93}, which is similar to the current value $1.01\times10^8{\rm~G}$ of J1946 \citep{bar17}.

\section{Simulated Results}

\begin{figure}
\centering
\includegraphics[width=0.75\linewidth,trim={30 210 220 30}]{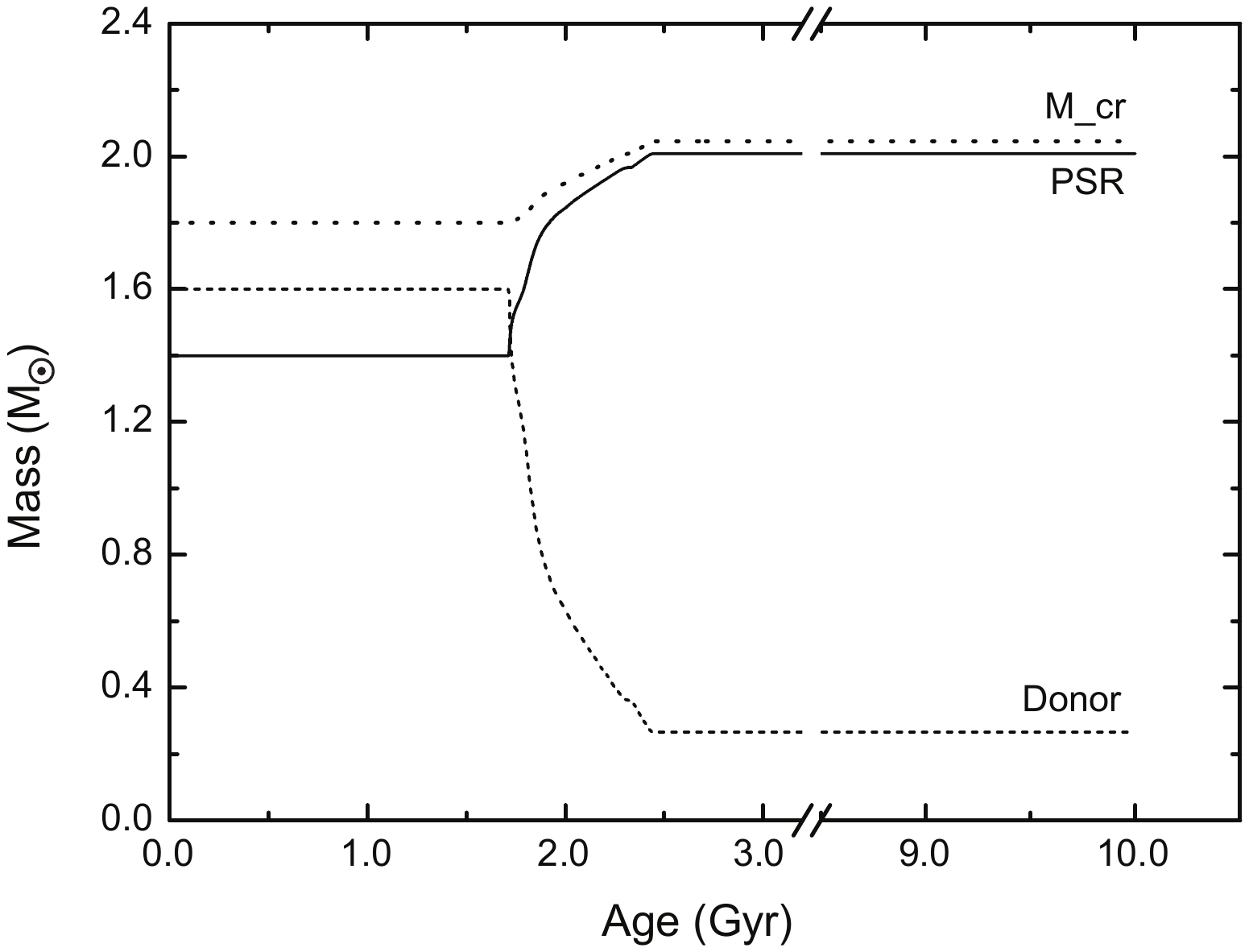}
\caption{\label{fig:per-m2} Mass evolution of Model A. The solid, dashed, and dotted curves represent the NS mass, the donor star mass, and the critical mass, respectively.
Since the mass of the NS is always lower than its critical mass, PT process is delayed to spin-down stage after RLOF ceases.}
\end{figure}

\begin{figure}
\centering
\includegraphics[width=0.75\linewidth,trim={30 210 220 30}]{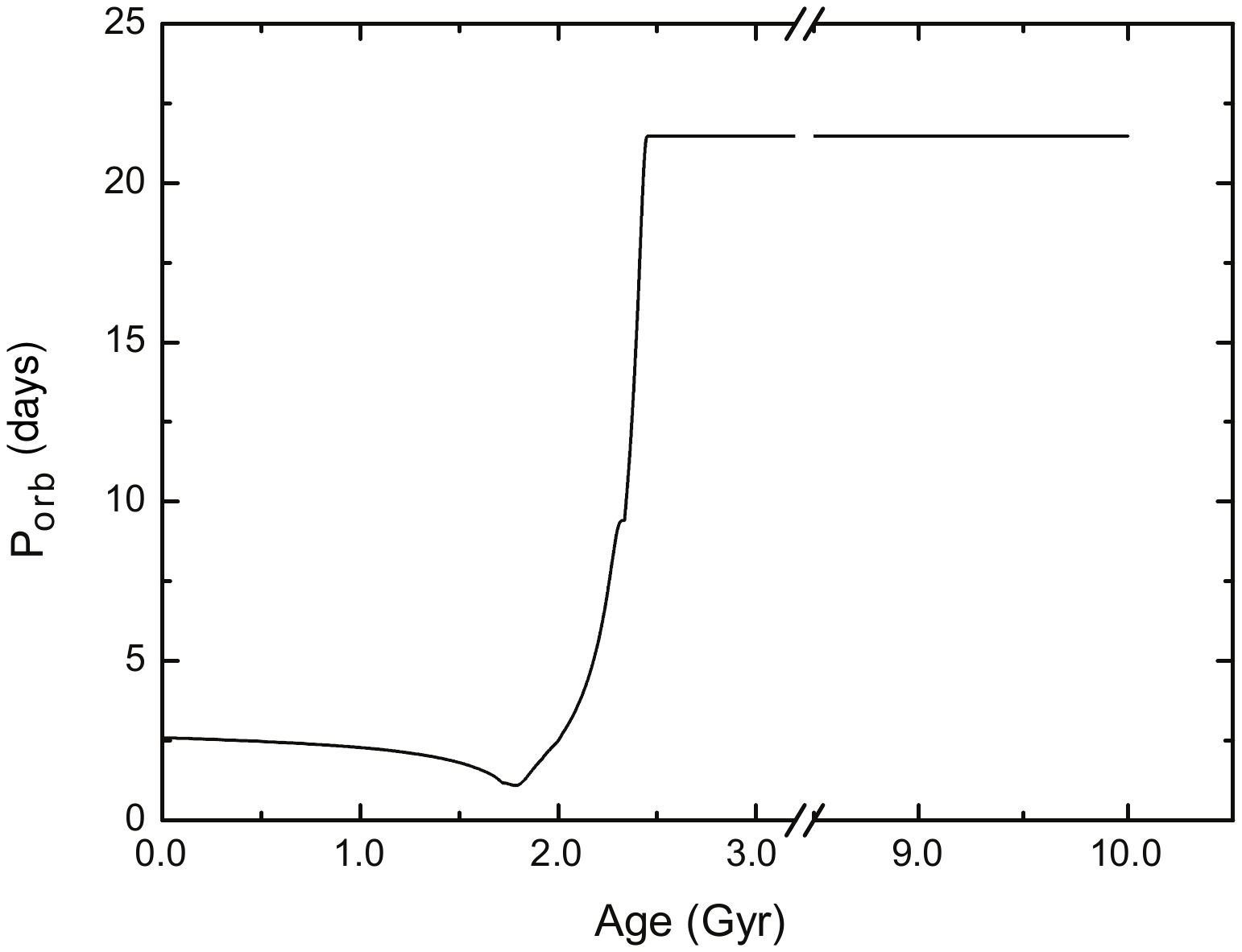}
\caption{\label{fig:per-m2} Orbital period evolution of Model A.}
\end{figure}

\begin{figure}
\centering
\includegraphics[width=0.6\linewidth,trim={60 20 80 40}]{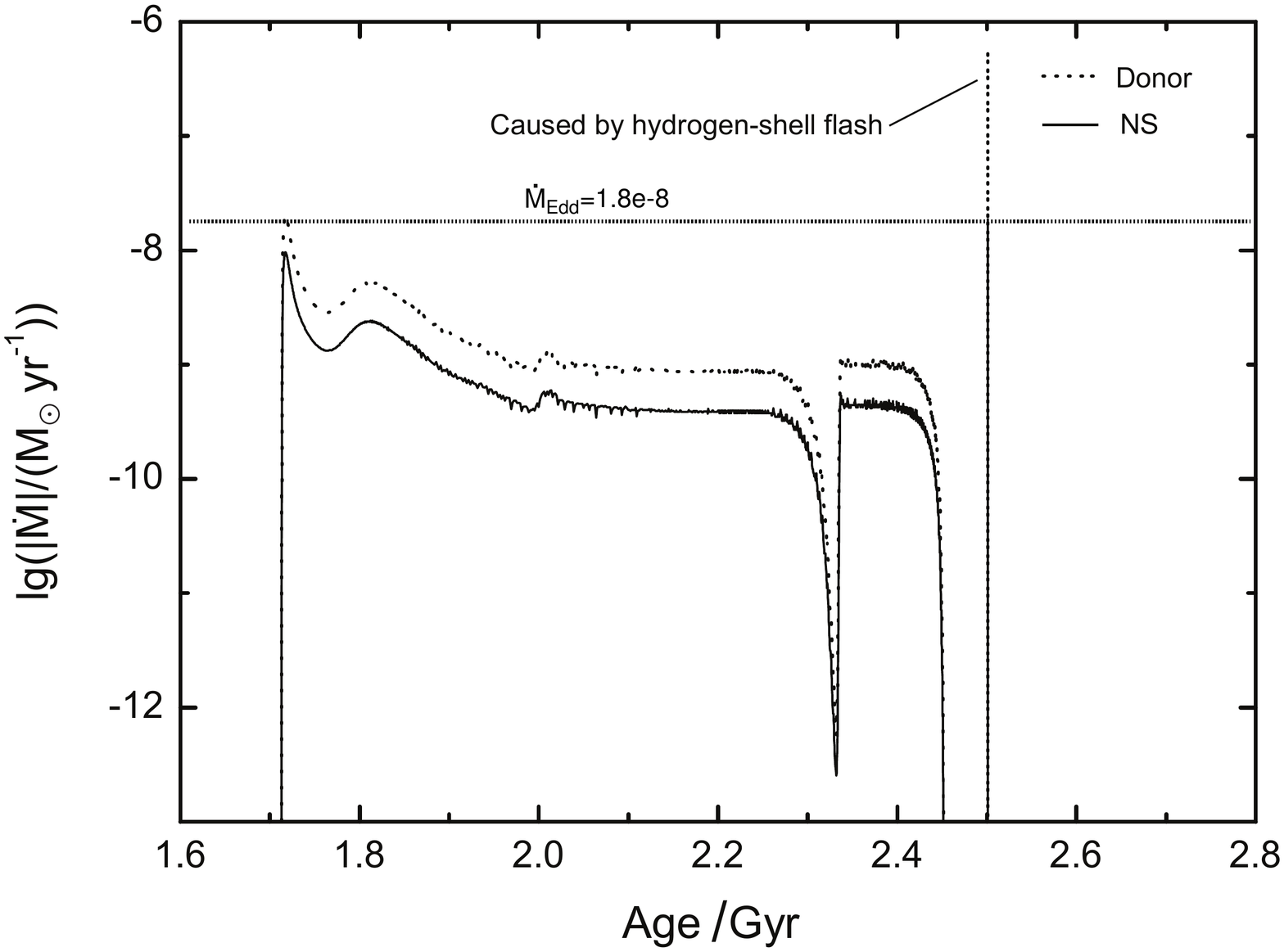}
\caption{\label{fig:per-m2} Evolution of the mass transfer rates. Dotted curve: the donor star. Solid curve: the NS.}
\end{figure}

\begin{figure}
\centering
\includegraphics[width=0.9\linewidth,trim={0 210 120 30}]{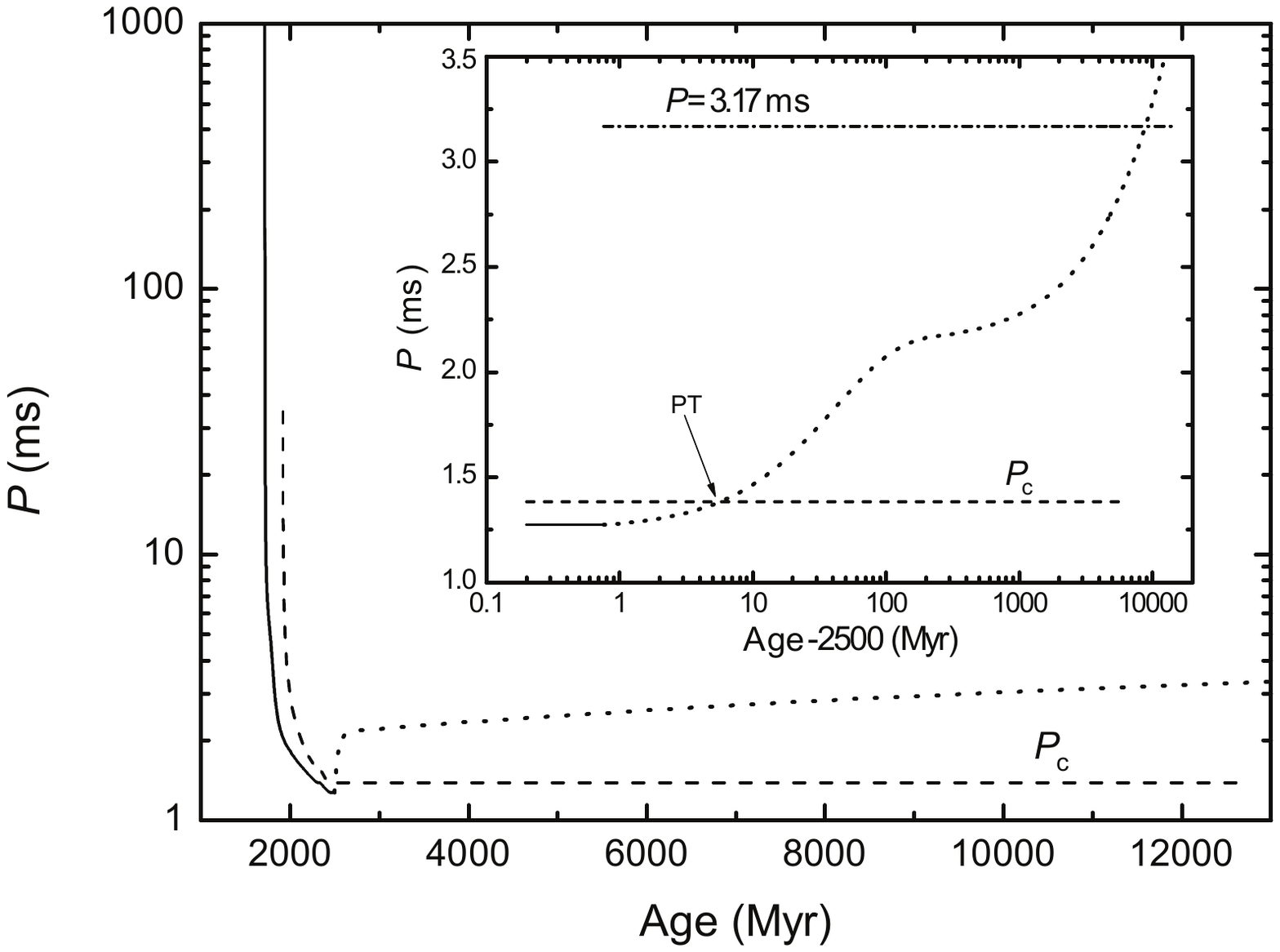}
\caption{\label{fig:pdot} Evolution of spin period of Model A. The solid, and dotted curves represent the spin-up (during RLOF) and the spin-down (mass transfer ceased) stages, respectively. The dashed curve denotes the evolution of critical period. In the small panel, the cross point between the dashed and the dotted curves corresponds to the moment that the PT takes place, and the dashed-dotted line indicates the current spin period of J1946.}
\end{figure}

\subsection{An evolutionary example}
With initial parameters, NS mass $M_{\rm NS, i}=1.4{\rm ~M}_\odot$, companion mass $M_{\rm d,i}=1.6{\rm ~M}_\odot$, orbital period $P_{\rm b,i}=2.59{\rm ~days}$, and $\beta=0.5$
\citep[which yields an accretion efficiency of 0.5,][]{pod02}, we evolve Model A as a typical example.
The masses evolutions of both the NS and the donor star are illustrated in Fig. 1.
The critical mass $M_{\rm c}(P)$ of NS-SS PT is denoted by the dashed curve.
It is clear that the NS mass is always lower than the critical mass during the mass transfer.
This implies that the PT event will be delayed to the spin-down stage of the NS after the mass transfer ceases.
Since the mass transfer timescale is as long as $\sim0.7{\rm ~Gyr}$,
the binary orbit will be circularized during RLOF, and keep circular until the PT takes place.
As a result, for NSs without PT, the binary MSPs will be observed in circular orbit.

Fig. 2 plots the orbital period evolution of Model A. Before the mass transfer or in the initial stage of the mass transfer, the orbit of pre-LMXB slowly shrink in a timescale of $\sim1.8$ Gyr due to the angular momentum loss driven by the magnetic braking. During the mass transfer ($\sim1.8-2.5~\rm Gyr$), the orbital period of LMXB gradually increase because the mass is transferring from the light donor star to the massive NS.
After RLOF ceases (at the stellar age $t_1\sim2.5{\rm ~Gyr}$), the orbital period is approximately constant since the angular momentum loss induced by gravitational wave radiation in such a wide binary is very weak.

Fig. 3 illustrates the evolution of the mass transfer rates.
Since the material is transferring from the massive donor star to the light NS, mass transfer occurs on the thermal timescale with a high rate ($\sim 10^{-8}~\rm M_{\odot}\,yr^{-1}$). Once the mass ratio is below 1, the mass-transfer rate will decreases to $\sim 10^{-9}~\rm M_{\odot}\,yr^{-1}$ sharply. The narrow single peak at the stellar age of $\sim2.5\rm ~Gyr$ is caused by the hydrogen-shell flash
while the upper edge of the solid line in it reached the Eddington accretion rate.
Since the flash may occur in most LMXB evolutions, formation of the CB disk and its effect (which is not considered in current work) requires more study.

Spin period evolution of Model A is illustrated in Fig. 4.
The solid and dotted curves represent the spin-up and spin-down evolutionary stages, respectively.
The dashed curve denotes the critical period that the PT occurs,
which can be derived by $M_{\rm c} (P)=M_{\rm NS}$ and Eq. (1).
In the small panel, the cross point between the dashed and the dotted curves indicates the moment that the PT occur,
which is at about $5{\rm ~Myr}$ (delay time, $t_2$) after the mass transfer ceases.
Furthermore the total spin-down timescale is $t_2+t_3\sim t_3\sim9{\rm ~Gyr}$.

\begin{table*}
\begin{center}
\footnotesize
\caption{Simulation for PSR J1946: some fine-tuned examples \label{tbl-2}}
\begin{tabular*}{0.65\textwidth}{@{}c|cccc|c@{}}
\hline\hline\noalign{\smallskip}
\footnotesize
                   & Model A & Model B & Model C & Model D     & J1946 \\\hline\noalign{\smallskip}
$\beta$      &  $0.5$          &  $0.5$     &  $0.3$       &   $0.5$                     & - \\
$M_{\rm NS,i}~({\rm M}_{\odot})$& $1.40$  & $1.35$ & $1.40$  & $1.60$      & -\\
$M_{\rm d,i}~({\rm M}_{\odot})$ & $1.60$    & $1.80$  & $1.30$  & $1.20$      & -\\
$P_{\rm b,i}~({\rm days})$ & $2.59$  & $2.46$ &$2.87$   & $2.77$       & -\\
$M_{\rm d,f}~({\rm M}_{\odot})$ & 0.2651  & 0.2643 & 0.2651  & 0.2656       &0.2656(19)\\
$t_1~(\rm Gyr)$ &  2.5   &  1.7 & 4.6 & 6.3     & -   \\
$P_{\rm b,f}~({\rm days})$& 21.47  & 20.83 & 21.64  & 22.10  & 27.02\\
$L_{\rm WD}~(\rm L_{\odot}/1000)$ &0.23        &0.16   &0.67       &1.15         &-\\
$T_{\rm WD, eff}~(\rm 1000K)$        &5.0           &4.6     &6.3            &7.2                 &-\\
$M_{\rm NS,f}~({\rm M}_{\odot})$& $2.01$  & $2.05$ &$2.06$   & $2.03$     & -\\
$M_{\rm c,f}~({\rm M}_{\odot})$ & 2.05  & 2.13 & 2.10 &  1.93     & - \\
Delayed PT & Yes & Yes & Yes  & No  & -\\
$M_{\rm SS}~({\rm M}_{\odot})$& 1.809  & 1.845 &1.854   & -    & 1.828(22)\\
$P_{0}~({\rm ms})$ &1.274   &  1.106  & 1.164 & -           & - \\
$P_{\rm c, f}~({\rm ms})$ & 1.383     & 1.277  & 1.235   & -           & 3.17\\
$t_2~(\rm Myr)$ &  4.9    & 7.1   & 2.8 & -        &- \\
$t_3~(\rm Gyr)$ & 8.9       &   9.5   &9.3 & -           & -\\
$t_T~(\rm Gyr)$ & 11.4    &   11.2   &13.9 & -        & -\\
\hline\noalign{\smallskip}
\end{tabular*}
\end{center}
\end{table*}

\subsection{Selected Models}
The input parameters and the simulated results of some models are summarized in Table 2.
Their initial parameters, $M_{\rm NS, i}$, $M_{\rm d,i}$ and $P_{\rm b,i}$ are selected to evolve pre-LMXB to the progenitor of J1946 at the endpoint of RLOF, which is characterized by following points:
(1) the final mass of the NS is $2.00\rm~M_{\odot}\leq M_{\rm NS,f}\leq2.06\rm~M_{\odot}$; (2) the companion is a He WD with a mass of $0.2640\leq M_{\rm d,f}\leq0.2680\rm~M_{\odot}$ \citep{bar17}.
According to the relation between the WD mass and the orbital period \citep{tau99},
all models predict a narrow final orbital period range: $20.8\leq P_{\rm b, f}\leq 22.1{\rm ~days}$.

As shown in Table 2, the NS is spun up to $P_{0}$ at the endpoint of RLOF.
Inserting $P_{0}$ into Eq.(1), one can derive the critical mass $M_{\rm c,f}$.
Comparing with the final mass of the NS, we find that, similar to Model A, Models B \& C are expected delayed PT models since $M_{\rm c,f}>M_{\rm NS, f}$. The PT will occur when the NSs spin down to the critical spin period, $P_{\rm c,f}$,
which is derived by $M_{\rm c}(P)=M_{\rm NS, f}$ and Eq. (1).
For Model D, the NS-SS PT will take place during its RLOF stage.
In the following evolution, the SS accretes a mass $\geq0.1\rm ~M_\odot$ in a timescale of $\geq10^7\rm ~yr$ and its orbit is re-circularized. Since delayed PT models are much rare (result of Paper 2), the results of NS-SS PT scenario are consistent with the observation that most MSPs are harbored in circular binaries.

For the PT delayed models, since the delay time, $t_2$,  is only several Myr, the total spin-down time is $t_2+t_3\sim t_3\sim9\rm~Gyr$.
Furthermore, the total evolutionary time, $t_{\rm T}=t_{1}+t_2+t_{3}$, of these models approximate to $11.4\rm~Gyr$ except Model C. It seems that Model C should be ruled out since its total evolutionary timescale is slightly longer than the Hubble time (13.7$\rm~Gyr$). However, a shorter spin-down timescale can be obtained with some fine-tuned period evolution parameters, e.g. a longer $t_{\rm D}$.

To compare with the observation, some observed parameters of J1946 are also listed in Table 2.
It is worth noting that, since (1) the NS-SS PT process is not included in the \texttt{MESA},
and (2) $P_{\rm c,f}$ is the critical period that the PT occurs,
$P_{\rm b,f}$ and $P_{\rm c,f}$ can not be compared directly with
the orbital period and spin period of J1946.
In addition, the luminosities and effective temperatures of the WDs (at the stellar age about $13{\rm ~Gyr}$)
predicted by the \texttt{MESA} simulation are also given for further checking when more observations are available.

\section{PT kick}

\begin{figure}
\centering
\includegraphics[width=0.8\linewidth,trim={40 10 50 40}]{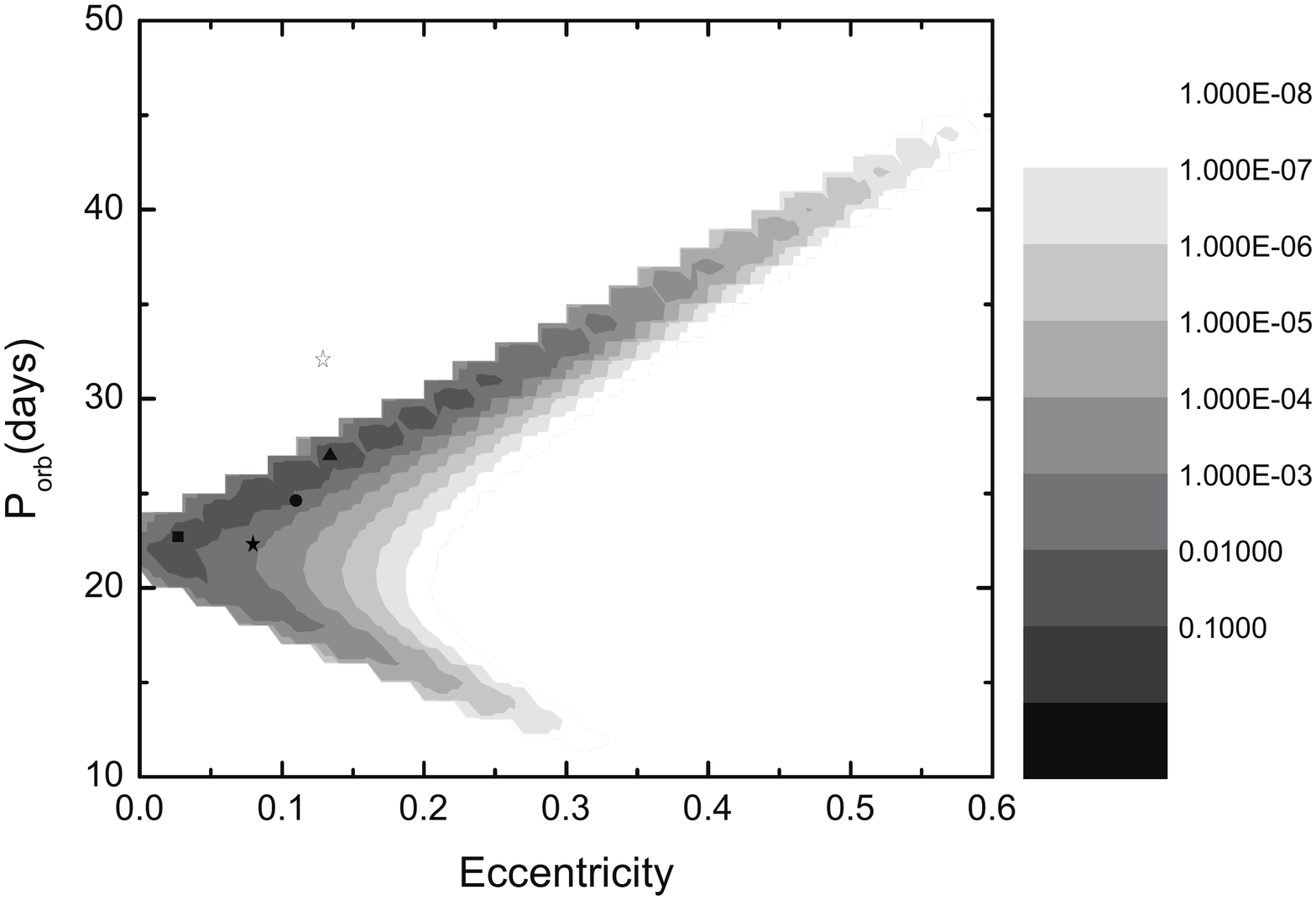}
\caption{\label{fig:per-m2} Possibility distribution on the $P_{\rm b}-e$ diagram after PT kick.
The various level of gray represents the possibility distribution. The solid triangle, square, filled star, empty star and circle represent J1946, J1618, J1950, J2234, and J0955, respectively.}
\end{figure}

Taking the simulated results of \texttt{MESA}: $M_{\rm NS}=2.03{\rm~M_{\odot}}$, $M_{\rm d}=0.265{\rm~M_{\odot}}$ and $P_{\rm b}=21.5$ days as input parameters before PT,
we simulated the influence of PT on the post-LMXB by the subprogram \textsf{kick.f} of \texttt{BPS} developed by \citet{hur00, hur02}. A weak kick with velocity of $\sigma_{\rm PT} = 60\rm~km~s^{-1}$ \citep{hob05},
and gravitational mass loss $\Delta M=0.1M_{\rm NS}$ during PT are adopted following Paper 2.
The probability distribution of $10^{10}$ PT events in the $P_{\rm b}-e$ diagram is shown in Fig. 4.
The total probability that PT processes producing binary MSPs with $0.12\leq e\leq 0.14$,
and $26\leq P_{\rm b}\leq 28$ is about $7.2\%$. Therefore, the NS-SS PT process is most likely evolutionary channel to J1946.

In Fig. 5, the solid triangle, square, and circle correspond to J1946, J1618, and J0955, respectively.
Although there is no further mass measurement of J1618 and J0955 except for the mass functions,
the $P_{\rm b}-e$ diagram show that the NS-SS PT process may be responsible for the formation of these two sources.
The another two eMSPs J1950 \& J2234 are also shown as filled and open stars in the figure.
However, their measured masses are too low to consistent with the NS-SS PT scenario. 

Furthermore, the effects of PT on the post-LMXB include two respects.
The first one is sudden mass loss, which results in an orbital expansion (the orbital period after PT is longer than before PT) and a constant eccentricity $e=\Delta M/M_{\rm T}$, as discussed in Paper 1.
The second one is the kick (see also Eq. 4), which results in spread distribution (alter the orbital period and eccentricity) as shown in Fig. 5. The probability of PT producing orbital periods greater and less than $21.5 \rm ~days$ are $88.6\%$, and $11.4\%$, respectively.
In addition, the probability resulting in larger eccentricity ($e>0.2$) is relatively low ($\sim15\%$).
Obviously, a low or high kick velocity will result in a narrow or wide distribution range in the orbital period versus eccentricity diagram, respectively.

\section{Summary}

In this work, we propose a delayed NS-SS PT scenario to account for the formation of eMSP J1946.
Employing the stellar evolution code \texttt{MESA}, we simulated the evolution of its progenitor.
The calculations indicate that a pre-LMXB consisting of a NS with an initial mass of $1.35\leq M_{\rm NS,i}\leq1.4\rm~M_\odot$ and a MS companion star (with a fine-tuning initial mass and an initial orbital period) can evolve into a post-LMXB consisting of a $\sim2.0\rm~M_\odot$ NS and a $\sim0.27\rm~M_\odot$ WD in an orbit of $\sim22$ days. Because of a rapidly rotation, the NS would not collapse to SS during the RLOF. After the mass transfer ceases, NS-SS PT will take place when the NS spin down to a critical spin period $P_{\rm c, f}$.

Based on the simulated results of \texttt{MESA}, we study the influence of PT on the post-LMXB.
A simulation of $10^{10}$ PT events is performed via the subprogram \textsf{kick.f} of \texttt{BSE} with a gravitational mass loss $\Delta M=0.1~M_{\rm NS}$, and weak kick $\sigma_{\rm PT}=60 ~{\rm ~km~s}^{-1}$.
The result shows that there are $\geq7\%$ PT events producing $26\leq P_{\rm b} \leq 28 \rm ~days$
and $0.12\leq e\leq0.14$, which are in good agreement with J1946.
Our simulations also predict luminosities and effective temperatures of the WDs,
which might be compared with and testified by future observations. 

Additionally, the positions of J1618 and J0955 on the $P_{\rm b}-e$ diagram imply that they probably experienced evolutionary processes similar to J1946 which will be studied in detail when more mass informations are available.

\normalem
\begin{acknowledgements}
We thank the referee for the comments that have led to the improvement of the manuscript. We thank Xiang-Dong Li and Zhi-Fu Gao for their helpful discussion. This work was supported by the CAS 'Light of West China' Program (Grants No. 2018-XBQNXZ-B-022), the National Natural Science Foundation of China (Grant Nos. 11803018, 11733009, 11773015, U2031116, and 11605110), the National Key Research and Development Program of China (Grant No. 2016YFA0400803), the Program for Innovative Research Team (in Science and Technology) at the University of Henan Province and the Key laboratory of Modern Astronomy and Astrophysics (Nanjing University), Ministry of Education.
\end{acknowledgements}


\end{document}